\documentclass[aps,prd,twocolumn,groupedaddress,showpacs,showkeys]{revtex4-1}
\usepackage{graphicx}
\usepackage{dcolumn}
\usepackage{bm}
\usepackage{amssymb,amsmath,latexsym,amsfonts}
\begin{document}


\title{Ultra--cold gases and the detection of the Earth's rotation:
Bogoliubov space and gravitomagnetism}

\author{A. Camacho}
 \email{acq@xanum.uam.mx} \affiliation{Departamento de F\'{\i}sica,
 Universidad Aut\'onoma Metropolitana--Iztapalapa\\
 Apartado Postal 55--534, C.P. 09340, M\'exico, D.F., M\'exico.}

\author{E. Castellanos}
\email{elias.castellanos@zarm.uni-bremen.de}\affiliation{ZARM,
Universit\"at Bremen, Am Fallturm, 28359 Bremen, Germany}

 \date{\today}

 \begin{abstract}
The present work analyzes the consequences of the gravitomagnetic
effect of the Earth upon a bosonic gas in which the corresponding
atoms have a non--vanishing orbital angular momentum. Concerning the
ground state of the Bogoliubov space of this system we deduce the
consequences, on the pressure and on the speed of sound, of the
gravitomagnetic effect. We prove that the effect on a single atom is
very small, but we also show that for some thermodynamical
properties the consequences scale as a non--trivial function of the
number of particles.
\end{abstract}

 \maketitle

\date{\today}

The deep analogies that can be found between electromagnetism and
gravitation have a long history, a fact readily understood looking
at Coulomb's law of electricity and Newton's law of gravitation
\cite{Ciufolini1}. The analogy was taken further and the possible
existence of a {\it magnetic} component in the gravitational
interaction between the Sun and the remaining celestial objects of
our solar system was put forward in the nineteenth century
\cite{Holz, Tisser}. A more complete and profound relation between
these two interactions emerged with the formulation of Einstein's
theory of general relativity (GR), in which a gravitomagnetic field
appears as an inexorable consequence of the presence of a current of
mass--energy \cite{Ciufolini1}, though several theories of
gravitation predict a gravitomagnetic contribution \cite{Ni}. This
effect has several consequences, one of them is related to the
appearance of a gravitational Larmor theorem \cite{Mash1}, i.e., the
exterior gravity of a rotating source couples to the angular
momentum of a test body and gives rise to a Larmor precession, in
the same way as a magnetic field couples with the angular momentum
of an electrically charged particle. Other manifestations of this
effect are the so--called frame--dragging and geodetic precession
\cite{Ciufolini2}

The gravitomagnetic field is one of the most important predictions
of GR and has no Newtonian counterpart. Currently the results
associated to the motion of the LAGEOS and LAGEOS--II satellites
provide observational evidence for this effect \cite{Ciufolini3}.
The extant spectrum of experiments, or of astrophysical
observations, have a weight, mainly, in the classical realm
\cite{Ciufolini2}. Of course, there are also experimental proposals
in the quantal world, namely, there is some evidence of a shift in
the energy for some fermions \cite{Venema, Mash2}. The need for a
more profound work in this direction is also a point that has to be
underlined, an issue already addressed in the context of
interferometry. either neutronic or atomic \cite{Mash3}. The quantal
aspect of the experimental corroboration has severe hurdles. Indeed,
this fact is related to the smallness of this effect (the ratio
between the gravitomagnetic and gravitoelectric effects of the Earth
is $\sim 10^{-7}$). Clearly, the use of an atomic system in this
context seems to be a very bad idea, due to the smallness of the
gravitational passive mass of an atom. This last assertion is a
correct one, though one must add that this statement is valid for
one atomic system. The point here concerns the possibility of
enhancing the consequences of this effect, upon a quantal system,
resorting to a gas. The present work addresses this issue, namely,
we show that in a Bose--Einstein condensate the effect upon a single
atom appears, in the context of some thermodynamical properties as
the speed of sound, multiplied by a non--trivial function of the
number of particles. This fact acts as an enhancer for this effect.

Let us consider a rotating uncharged, idealized spherical body with
mass $M$ and angular momentum $\vec {J}$. In the weak field and slow
motion limit the gravitomagnetic field may be written, using the PPN
parameters $\Delta_1$ and $\Delta_2$, as

\begin{equation}\vec {B}(\vec {x}) = \Bigl({7\Delta_1 + \Delta_2\over 4}\Bigr){G\over
c^2}{\vec {J} - 3(\vec {J}\cdot\hat {x})\hat {x}\over |\vec
{x}|^3}\label{Hamm3}.
\end{equation}

The case of GR implies ${7\Delta_1 + \Delta_2\over 4} = 1$
\cite{Mash1}, where $c$ is the speed of light, $\vec{x}$ is the
position vector, $G$ is the Newtonian gravitational constant, and
$\hat{x}$ is the unit vector related to $\vec{x}$ . We now assume
that this gravitomagnetic field couples to the orbital angular
momentum of an atom in the same way as it does in the case of a
classical angular momentum \cite{Mash1}.

Following this analogy between gravitomagnetism and magnetism we may
now write down the interaction Hamiltonian that describes the
coupling between gravitomagnetism and the orbital angular momentum
of an atom, here denoted by $\vec {L}$

\begin{equation}W = - \vec {L}\cdot\vec {B}\label{Hamm2}.
\end{equation}
\bigskip

For the sake of simplicity we now impose some restrictions. Firstly,
the Earth has a perfect spherical symmetry with a radius $R$, mass
$M$, constant mass density, and rotation frequency equal to
$\omega$. Under these conditions we have that \cite{Mash1}

\begin{equation}\vec {J} = \frac{2MR^2}{5}\omega\vec{e}_z\label{Hamm3}.
\end{equation}
\bigskip

Clearly, the Hamiltonian for an atom must include this last term the
one will be considered in the $N$--body Hamiltonian operator
(assuming that the gas is so dilute that only the two--body
interaction potential is required \cite{Pathria}). The system under
study will be a Bose--Einstein gas enclosed in a container of volume
$V$, particles of the gas are atoms with passive gravitational mass
$m$ and located at a height $z<<R$ with respect to the Earth's
surface. In addition, the interaction between two particles will be
assumed to be dominated by $s$--scattering, i.e., the temperature of
the system is very low ($ka<<1$, where $\vec{k}$ and $a$ are the
wave vector and the scattering length, respectively)
\cite{Pitaevski}. This entails the following Hamiltonian for the
$N$--body system.

\begin{eqnarray}
\hat{H} =
\sum_{\vec{k}=0}\frac{\hbar^2k^2}{2m}\hat{a}_{\vec{k}}^{\dagger}\hat{a}_{\vec{k}}\nonumber\\
+\frac{U_0}{2V}\sum_{\vec{k}=0}\sum_{\vec{p}=0}\sum_{\vec{q}=0}\hat{a}_{\vec{p}}^{\dagger}\hat{a}_{\vec{q}}^{\dagger}
 \hat{a}_{\vec{p}+\vec{k}}\hat{a}_{\vec{q}-\vec{k}} \nonumber\\
 +
 \sum_{\vec{k}=0}mgz\hat{a}_{\vec{k}}^{\dagger}\hat{a}_{\vec{k}} +
 \sum_{\vec{k}=0}\sum_{s=0, \pm 2}\frac{2sg\omega R\hbar}{5c^2}\hat{a}_{\vec{k}, s}^{\dagger}\hat{a}_{\vec{k},s}, \label{Ham1}
\end{eqnarray}

\begin{equation}
U_0=\frac{4\pi a\hbar^2}{m}.\label{Add1}
\end{equation}

The last term in our Hamiltonian is related to the fact that the
coupling between gravitomagnetism and orbital angular momentum is
absent for the case of vanishing $l$. In addition, since we assume
that the gas has a very low temperature (this phrase means smaller
than the condensation temperature) then almost all the particles
have $l=0$ and a few ones will have non--vanishing angular momentum
and, at this point, we assume that they are in the $d$--state, i.e.,
$l=2$, because the symmetry requirements associated to the wave
function discard the case of $l=1$ \cite{Pathria}, i.e., the first
case with non--vanishing angular momentum is $l=2$ and not $l=1$.
The parameter $s$ denotes the five possibilities related to the
eigenvalues of the operator $L_z$, namely $s=\pm 1, 0, \pm 2$. These
operators ($\hat{a}_{\vec{k}}$ and $\hat{a}_{\vec{k}}^{^{\dagger}}$)
are bosonic creation and annihilation operators, and fulfill the
usual Bose commutation relations. As an additional simplification,
which does not restrict the validity of our results, we assume that
all the particles in the $d$--state have eigenvalue for $L_z$ equal
to $s=+2$. Very close to the temperature $T=0$, the second term in
this Hamiltonian becomes \cite{Pethick}

\begin{eqnarray}
\sum_{\vec{k}=0}\sum_{\vec{p}=0}\sum_{\vec{q}=0}\hat{a}_{\vec{p}}^{\dagger}\hat{a}_{\vec{q}}^{\dagger}
 \hat{a}_{\vec{p}+\vec{k}}\hat{a}_{\vec{q}-\vec{k}} = N^2+2N\sum_{\vec{k}\not=0}\hat{a}_{\vec{k}}
 ^{\dagger}\hat{a}_{\vec{k}}\nonumber\\
 + N\sum_{\vec{k}\not=0}\Big(\hat{a}_{\vec{k}}^{\dagger}\hat{a}_{-\vec{k}}^{\dagger}
+ \hat{a}_{\vec{k}}\hat{a}_{-\vec{k}}\Big). \label{Add2}
\end{eqnarray}

With these approximations the $N$--body Hamiltonian has the
following structure

\begin{eqnarray}
\hat{H} = \frac{U_0N^2}{2V}+ m g z N+\nonumber\\
\sum_{\vec{k}\not=0}\Bigl[\frac{\hbar^2k^2}{2m}+ m g
z+\frac{4g\omega R\hbar}{5c^2}+\frac{U_0N}{V}\Bigr]\hat{a}_{\vec{k}}
^{\dagger}\hat{a}_{\vec{k}}\nonumber\\
+N\frac{U_0}{2V}\Bigl[\hat{a}_{\vec{k}}^{\dagger}\hat{a}_{-\vec{k}}^{\dagger}
+ \hat{a}_{\vec{k}}\hat{a}_{-\vec{k}}\Bigr]. \label{Ham2}
\end{eqnarray}

This Hamiltonian can be diagonalized introducing the Bogoliubov
transformations \cite{Pitaevski}

\begin{equation}
\hat{b}_{\vec{k}}=
\frac{1}{\sqrt{1-\alpha_k^2}}\Bigl[\hat{a}_{\vec{k}} +
\alpha_k\hat{a}_{-\vec{k}}^{\dagger}\Bigr],\label{Bog1}
\end{equation}

\begin{equation}
\hat{b}_{\vec{k}}^{\dagger}=
\frac{1}{\sqrt{1-\alpha_k^2}}\Bigl[\hat{a}_{\vec{k}}^{\dagger} +
\alpha_k\hat{a}_{-\vec{k}}\Bigr].\label{Bog2}
\end{equation}

These two operators fulfill the same algebra related to
$\hat{a}_{\vec{k}}$ and $\hat{a}_{\vec{k}}^{\dagger}$, i.e., they
are also bosonic operators. In this last expression the following
definitions have been introduced

\begin{equation} \epsilon_k =
\frac{\hbar^2k^2}{2m}+mgz+\frac{4g\omega R\hbar}{5c^2},\label{Add3}
\end{equation}

\begin{equation}
\alpha_k = 1+ \frac{V\epsilon_k}{U_0N}
-\sqrt{\frac{V\epsilon_k}{U_0N}}\sqrt{2+
\frac{V\epsilon_k}{U_0N}}.\label{Add33}
\end{equation}

The final form for our Hamiltonian is

\begin{eqnarray}
\hat{H} = \frac{U_0N^2}{2V} + mgzN \nonumber\\
+
\sum_{\vec{k}\not=0}\Bigl\{\sqrt{\epsilon_k(\epsilon_k+\frac{2U_0N}{V})}
\hat{b}_{\vec{k}}^{\dagger}\hat{b}_{\vec{k}}\nonumber\\
-\frac{1}{2}\Bigl[\frac{U_0N}{V} +\epsilon_k
-\sqrt{\epsilon_k(\epsilon_k+\frac{2U_0N}{V})}\Bigr]\Bigr\}.\label{Ham3}
\end{eqnarray}

The last summation diverges, a result already known \cite{Gribakin,
Ueda}, and this divergence disappears introducing the so--called
pseudo--potential method, which implies that we must perform the
following substitution \cite{Ueda}

\begin{eqnarray}
 &-\frac{1}{2}\Bigl[\frac{U_0N}{V} +\epsilon_k -\sqrt{\epsilon_k(\epsilon_k+\frac{2U_0N}{V})}\Bigr]&\rightarrow\nonumber \\
 &-\frac{1}{2}\Bigl[\frac{U_0N}{V}
 +\epsilon_k -\sqrt{\epsilon_k(\epsilon_k+
 \frac{2U_0N}{V})}-\frac{1}{2\epsilon_k}\bigl(\frac{U_0N}{V}\bigr)^2\Bigr]&.\label{Psedo1}
\end{eqnarray}

Finally, this last summation will be approximated by an integral. It
is noteworthy to mention that the original expression has as lower
limit the condition $k\not=0$, which implies that the integral does
not have as lower limit the value $0$. In other words,

\begin{eqnarray}
&-&\frac{1}{2}\sum_{\vec{k}\neq 0}\Bigl[\frac{U_0N}{V} +\epsilon_k
-\sqrt{\epsilon_k(\epsilon_k+
 \frac{2U_0N}{V})}-\frac{1}{2\epsilon_k}\bigl(\frac{U_0N}{V}\bigr)^2\Bigr]
 \nonumber\\
 &=& -\frac{\hbar^2V}{8m\pi^2}\bigl(\frac{8\pi aN}{V}\bigr)^{5/2}\int_{\alpha}^{\infty}f(x)dx,\label{Psedo2}
\end{eqnarray}

In this last expression we have that

\begin{eqnarray}
\alpha^2 =\Bigl[1+\frac{4\omega
R\hbar}{5mc^2z}\Bigr]\frac{mgzV}{U_0}.\label{Psedo33}
\end{eqnarray}
Additionally

\begin{eqnarray}
f(x) =
x^2\Bigl[1+x^2-x\sqrt{2+x^2}-\frac{1}{2x^2}\Bigr],\label{Psedo3}
\end{eqnarray}

\begin{eqnarray}
x = \sqrt{\frac{\epsilon_kV}{U_0N}}.\label{Psedo33}
\end{eqnarray}

With these conditions we deduce the final structure of the $N$--body
Hamiltonian

\begin{equation}
\hat{H} = E_0 +
\sum_{\vec{k}\not=0}E_k\hat{b}_{\vec{k}}^{\dagger}\hat{b}_{\vec{k}}.\label{Ham4}
\end{equation}

In this last expression $E_0$ denotes the energy of the ground state
of the corresponding Bogoliubov space \cite{Pitaevski}.

\begin{eqnarray}
E_0 = \frac{2\pi a\hbar^2N^2}{mV}\Bigl[1 +
\frac{128}{15}\sqrt{\frac{a^3N}{V\pi}}\nonumber\\
\Bigl(1-\frac{15} {16\sqrt{2}}\alpha\Bigr)\Bigr]
+\alpha^2\frac{U_0N}{V}. \label{E0}
\end{eqnarray}

On the other hand, we have that the energy of the Bogoliubov
excitations ($E_k$) is given by \cite{Pitaevski}

\begin{equation}
E_k = \sqrt{\epsilon_k\Bigl(\epsilon_k+
 \frac{2U_0N}{V}\Bigr)}. \label{Ek}
\end{equation}

Concerning (\ref{Ham4}), if we impose the condition of vanishing
gravitational constant, i.e. $g=0$, then we recover the usual
Hamiltonian \cite{Ueda}.
\bigskip

The pressure ($P_0 = -\frac{\partial E_0}{\partial V}$) and speed of
sound ($v_s =\sqrt{-\frac{{V}^{2}}{N m}\frac{\partial P_0}{\partial
V}}$) associated to the ground state of the Bogoliubov space become,
respectively

\begin{eqnarray}
P_0 = \frac{2\pi a\hbar^2N^2}{mV^2}\Bigl[1 +
\frac{194}{15}\sqrt{\frac{a^3N}{V\pi}}\nonumber\\
\Bigl(1-\frac{15} {16\sqrt{2}}\alpha\Bigr)\Bigr]+ O(\alpha^2),
\label{Pre0}
\end{eqnarray}

\begin{eqnarray}
v^2_s = \frac{4\pi a\hbar^2N}{m^2V}\Bigl[1 +
\frac{242}{15}\sqrt{\frac{a^3N}{V\pi}}\nonumber\\
\Bigl(1-\frac{15}{16\sqrt{2}}\alpha\Bigr)\Bigr]+ O(\alpha^2).
\label{Sp0}
\end{eqnarray}

Notice that the possibility of detecting the term depending upon the
gravitomagnetic effect ($\delta v^{gm}_s$) requires that, if $\Delta
v_s$ is the experimental error related to the measurement of the
speed of sound, then $\Delta v_s <\vert\delta v^{gm}_s\vert$. In our
case this entails

\begin{eqnarray}
\Delta(v_s) < \frac{3}{\sqrt{2}}\frac{a\omega gR}{c^2}N.\label{Sp10}
\end{eqnarray}

The detection of the speed of sound in condensates has already a
long history \cite{Andrews, Andrews2}. The main difficulty, in the
experimental context, is related to the fact that (for the case of
an atom) $\frac{a\omega gR}{c^2}\sim 10^{-23}$. Nevertheless, this
contribution to the speed of sound does not depend upon the density
of particles but upon the number of particles. In other words, this
tiny contribution is enhanced by the number of particles ($N$)
related to a bosonic gas. The density in condensation experiments
ranges from $10^{13}$ to $10^{15}$ particles per cubic cm
\cite{Griffin}, hence for a condensate whose volume is $1.4cm^3$ we
have that

\begin{eqnarray}
\frac{3}{\sqrt{2}}\frac{a\omega gR}{c^2}N\sim
10^{-2}m/s.\label{Sp11}
\end{eqnarray}

Under these conditions an experimental uncertainty of $\Delta
v_s\sim 10^{-3}m/s$ would allow the detection of this field.

\begin{acknowledgments}
This research was partially supported by DAAD grant $A/09/77687$
\end{acknowledgments}

\end{document}